\documentclass[11pt]{article}

\usepackage[preprint]{acl}

\usepackage{times}
\usepackage{latexsym}
\usepackage[T1]{fontenc}
\usepackage[utf8]{inputenc}
\usepackage{microtype}
\usepackage{inconsolata}

\usepackage{graphicx}
\usepackage{booktabs}
\usepackage{multirow}
\usepackage{amsmath}
\usepackage{amssymb}
\usepackage{algorithm}
\usepackage{algorithmic}
\usepackage{subcaption}
\usepackage{enumitem}
\setlist{nosep,leftmargin=*}
\setlist[enumerate]{label=\arabic*.}

\usepackage{xcolor}
\newcommand{\rev}[1]{#1}  

\usepackage{caption}
\title{The Reasoning Bottleneck in Graph-RAG: \\
Structured Prompting and Context Compression for Multi-Hop QA}

\author{
  Yasaman Zarrinkia \\
  University of Victoria \\
  \texttt{yasamanz@uvic.ca} \\\And
  Venkatesh Srinivasan \\
  Santa Clara University \\
  \texttt{vsrinivasan4@scu.edu} \\\And
  Alex Thomo \\
  University of Victoria \\
  \texttt{thomo@uvic.ca} \\
}

\begin{document}
\maketitle

\begin{abstract}
Graph-RAG systems achieve strong multi-hop question answering by indexing documents into knowledge graphs, but strong retrieval does not guarantee strong answers.
Evaluating KET-RAG, a leading Graph-RAG system, on three multi-hop QA benchmarks (HotpotQA, MuSiQue, 2WikiMultiHopQA), we find that 77\% to 91\% of questions have the gold answer in the retrieved context, yet accuracy is only 35\% to 78\%, and 73\% to 84\% of errors are reasoning failures.
We propose two augmentations:
(i)~SPARQL chain-of-thought prompting, which decomposes questions into triple-pattern queries aligned with the entity-relationship context, and
(ii)~graph-walk compression, which compresses the context by ${\sim}$60\% via knowledge-graph traversal with no LLM calls.
SPARQL CoT improves accuracy by +2 to +14~pp; graph-walk compression adds +6~pp on average when paired with structured prompting on smaller models.
Surprisingly, we show that, with question-type routing, a fully augmented budget open-weight Llama-8B model matches or exceeds the unaugmented Llama-70B baseline on \emph{all three benchmarks} at ${\sim}12\times$ lower cost.
A replication on LightRAG confirms that our augmentations transfer across Graph-RAG systems.
\end{abstract}

\section{Introduction}

Multi-hop question answering (QA) requires reasoning over multiple pieces of evidence scattered across documents.
Graph-based retrieval-augmented generation (Graph-RAG) addresses this by constructing knowledge graphs from document corpora and using graph structure to retrieve relevant context~\cite{graphrag2024,ketrag2025}.

\paragraph{The retrieval-reasoning gap.}
Strong retrieval does not guarantee strong answers.  Using KET-RAG~\cite{ketrag2025}, a state-of-the-art Graph-RAG system, with budget open-weight models (\S\ref{sec:setup}), we achieve 77\% to 91\% context coverage across three benchmarks (Table~\ref{tab:coverage}).
Yet overall accuracy ranges from only 23\% to 67\% for Llama-3.1-8B and 35\% to 78\% for Llama-3.3-70B.
The culprit is context volume: KET-RAG's retrieved context typically spans ${\sim}$10{,}000 tokens, containing entity descriptions, relationship descriptions, community report summaries, and raw text chunks.
The model must locate the relevant facts and chain them together within this large context, a needle-in-a-haystack challenge that intensifies with the number of reasoning hops required.
Decomposing errors into retrieval failures (answer absent from context) and reasoning failures (answer present but model incorrect) reveals that reasoning failures account for 73\% to 84\% of all errors across datasets.
This raises a natural question: can inference-time augmentations to the reasoning stage, requiring no retraining or re-indexing, close this gap?

\paragraph{Reasoning augmentations.}
To address this bottleneck, we propose two synergistic strategies that require no additional model training.
Our augmentations apply to any Graph-RAG system whose retrieved context preserves entity-relationship structure; we focus on KET-RAG and verify that the strategies transfer to LightRAG~\cite{lightrag2024}, another prominent Graph-RAG system.

\begin{enumerate}
    \item \textbf{SPARQL CoT prompting} (\S\ref{sec:sparql_cot}): A structured chain-of-thought prompt that asks the model to decompose the question as a SPARQL query over the retrieved facts, leveraging the inherent entity-relationship structure of the knowledge-graph context.
    This is a single LLM call with a modified prompt; the only additional cost is slightly longer output.
    \item \textbf{Graph-walk context compression} (\S\ref{sec:gw}): A post-retrieval filtering step that performs breadth-first traversal from question entities through the knowledge graph, retaining only structurally connected context.
    This reduces input tokens by ${\sim}$60\% with no LLM calls, directly reducing the needle-in-a-haystack challenge.
\end{enumerate}

Evaluated on three standard multi-hop QA benchmarks~\cite{hotpotqa2018,musique2022,2wikimhqa2020} across multiple model sizes and configurations (\S\ref{sec:setup}), our contributions are:
\begin{enumerate}
    \item A quantitative error decomposition showing that reasoning, not retrieval, is the dominant bottleneck: 73\% to 84\% of errors occur when the gold answer \emph{is} present in the retrieved context.
    \item SPARQL CoT \rev{consistently improves over the no-CoT baseline (+2 to +14~pp)}, with the largest gains on harder datasets; an ablation (\S\ref{sec:cot_ablation}) confirms that \rev{structured decomposition is the primary driver, with SPARQL particularly effective on 70B}.
    \item A systematic analysis revealing a synergy between graph-walk compression and structured prompting: compression most helps when the model is following a reasoning roadmap, with smaller models gaining +6~pp average.
    \item The finding that combining structured prompting, graph-walk compression, and question-type routing enables a budget 8B model to match or exceed the unaugmented 70B baseline on \emph{all three benchmarks} at ${\sim}12\times$ lower cost.
    \item A generality check showing that SPARQL CoT transfers to LightRAG (+7.6~pp on 2WikiMHQA), while graph-walk compression's effectiveness depends on retrieval-pipeline structure rather than context size alone.
\end{enumerate}

\section{Background: KET-RAG}
\label{sec:background}

KET-RAG~\cite{ketrag2025} is a multi-granular indexing framework built on Microsoft's GraphRAG~\cite{graphrag2024}.
It reduces the cost of full-corpus entity extraction by combining three retrieval strategies (\textbf{K-E-T}):
(K)~a text-keyword bipartite graph built from all chunks without LLM calls;
(E)~LLM-based entity/relationship extraction on a PageRank-selected subset of core chunks ($\beta{=}0.8$), producing a knowledge graph queried via GraphRAG's local search;
and (T)~standard text-chunk retrieval via embedding similarity.
A parameter $\theta \in [0, 1]$ balances the entity and keyword channels within a total budget of $\lambda$ tokens; we use $\theta{=}0.5$, $\lambda{=}12{,}000$ (defaults of the published code).

KET-RAG achieves state-of-the-art results on multi-hop QA, outperforming all competing systems in its evaluation~\cite{rag2020,knngrag2024,graphrag2024,hybridrag2024,hyde2023,hipporag2024,lightrag2024}.
More broadly, KET-RAG is an instance of a growing family of Graph-RAG systems, including LightRAG~\cite{lightrag2024} and GraphRAG~\cite{graphrag2024}, whose retrieved context preserves explicit entity-relationship structure (entity descriptions, relationship triples, community summaries).
This structure creates an opportunity: rather than treating the context as flat text, reasoning augmentations can exploit its entity-relationship layout.
We validate this generality on LightRAG in Appendix~\ref{app:lightrag}.
We replace KET-RAG's commercial components (GPT-4o-mini, OpenAI embeddings) with budget open-weight alternatives without loss in coverage (\S\ref{sec:coverage}) and focus on closing the reasoning gap.

\section{Methods}
\label{sec:methods}

\subsection{SPARQL Chain-of-Thought Prompting}
\label{sec:sparql_cot}

\begin{figure}[t]
\centering
\includegraphics[width=\columnwidth]{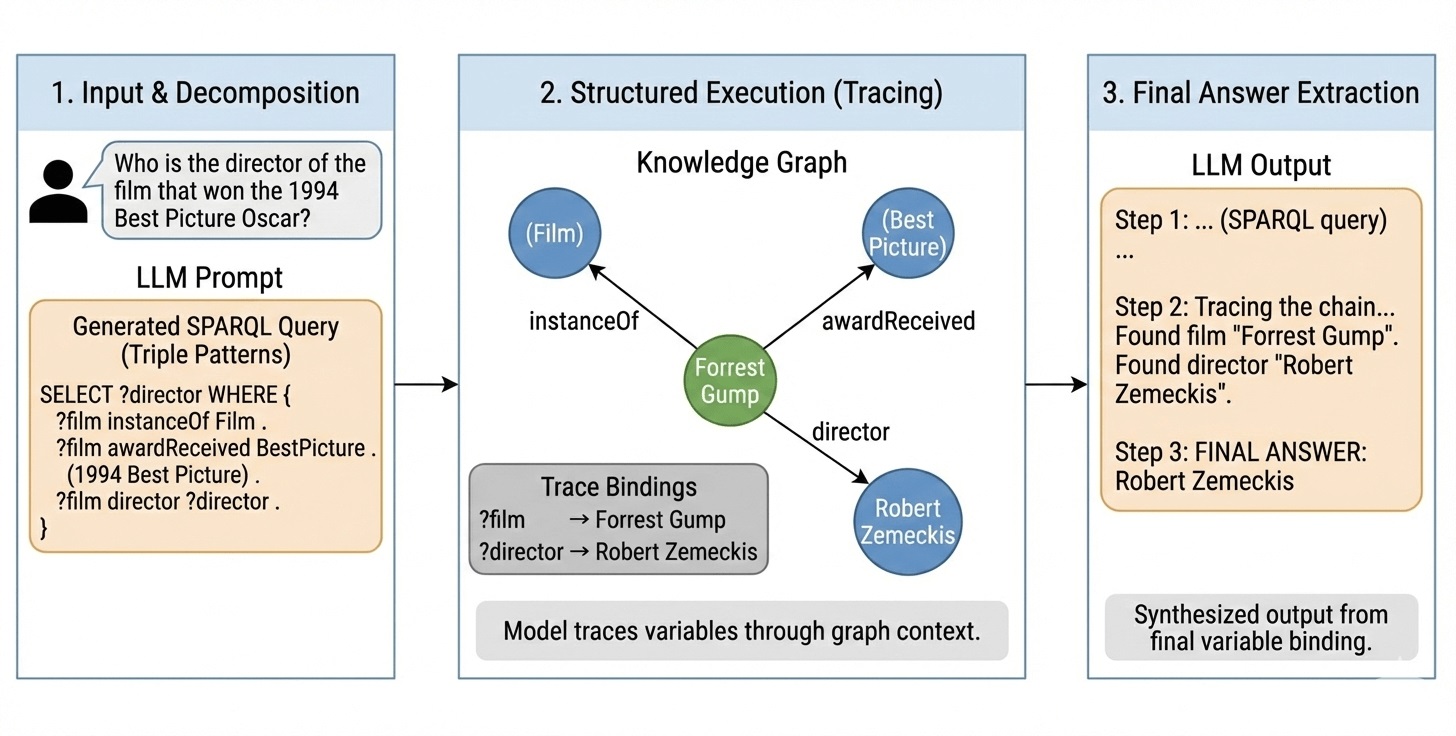}
\caption{SPARQL CoT prompting.  The model generates a SPARQL query whose triple patterns align with the retrieved context, then traces variable bindings to produce the answer in a single LLM call.}
\label{fig:sparql_cot}
\end{figure}

Multi-hop questions require the model to identify and chain multiple facts scattered across a large retrieved context.
Standard prompting leaves the model to perform this chaining implicitly, often causing it to miss intermediate hops or abstain when the reasoning path is unclear.

We observe that KET-RAG's retrieved context is inherently structured: it consists of \emph{entity descriptions}, \emph{relationship descriptions} (subject-predicate-object triples), \emph{community report summaries}, and \emph{text chunks}.
This structure closely mirrors a knowledge graph that can be queried with a formal language.
We exploit this alignment by introducing a chain-of-thought (CoT) prompting strategy that uses SPARQL~\cite{sparql2013}, the W3C standard query language for RDF knowledge graphs, as a reasoning formalism.
SPARQL's triple-pattern syntax (\texttt{?subject predicate ?object}) naturally expresses multi-hop entity chains, making it a good fit for decomposing questions over entity-relationship context.
Rather than asking the model to ``think step by step'' in natural language, we ask it to formulate the question as a structured graph query whose triple patterns map directly onto the entity-relationship pairs in the context (Figure~\ref{fig:sparql_cot}).

The prompt template is:

\begin{quote}
\small
\texttt{Step 1:} Write a simple SPARQL query (max 4 triple patterns, plain English predicates, NO URIs, NO FILTER, NO subqueries). Example: \\
\texttt{~~SELECT ?answer WHERE \{} \\
\texttt{~~~~?x name "Paradise Creek" .} \\
\texttt{~~~~?x tributaryOf ?y .} \\
\texttt{~~~~?y tributaryOf ?answer .} \\
\texttt{~~\}} \\
\texttt{Step 2:} Trace the variable bindings for each triple pattern through the context. \\
\texttt{Step 3:} Write \texttt{FINAL ANSWER: <your answer>} \\
\texttt{If the answer is not in the context, write FINAL ANSWER: I don't know}
\end{quote}

This is a \emph{single} LLM call: the model generates the SPARQL query, traces it through the context, and produces the answer in one response.
The key advantage over generic sub-question decomposition is \emph{structural alignment}: the triple patterns (e.g., \texttt{?x tributaryOf ?y}) correspond directly to the relationship descriptions in the retrieved context, converting an open-ended search over ${\sim}$10{,}000 tokens into a structured template-matching task where each variable binding narrows the search space for the next hop.
The only additional cost is the longer output (the SPARQL query and step-by-step trace, typically 100--200 extra tokens).

The final answer is extracted via regex matching on \texttt{FINAL ANSWER:} in the model's response.  If the model determines the answer is not in the context, it may respond with ``I don't know,'' which we count as an abstention.

\subsection{Graph-Walk Context Compression}
\label{sec:gw}

While KET-RAG achieves high coverage, much of its ${\sim}$10{,}000-token context is topologically distant from the question's reasoning chain, increasing the needle-in-a-haystack difficulty.

We propose \textbf{graph-walk context compression (GW)}, a post-retrieval step that exploits the knowledge graph's own topology to isolate a compact \emph{reasoning subgraph} relevant to the query, with no additional LLM or embedding costs.
The algorithm distills the context in four phases:

\begin{figure}[htbp]
\centering
\includegraphics[width=\columnwidth]{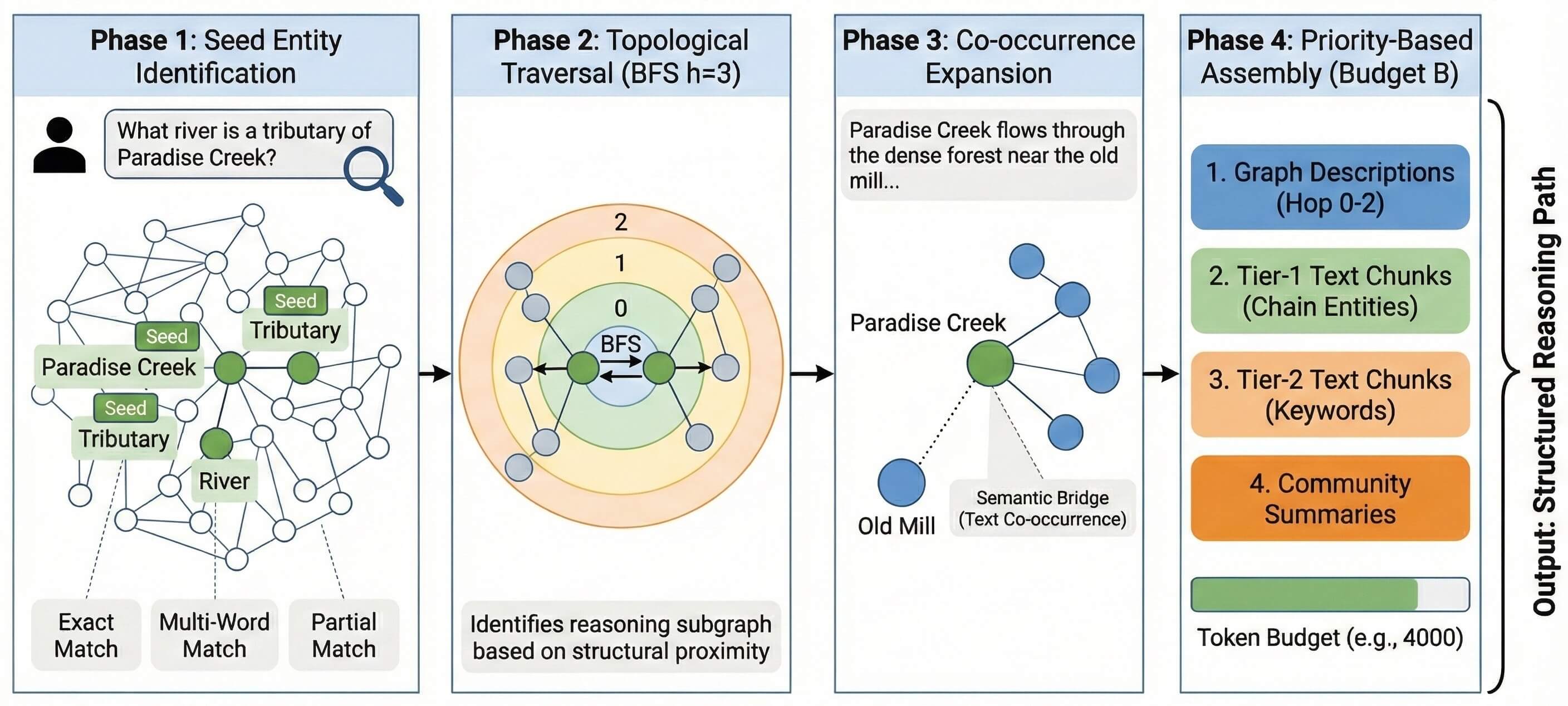}
\caption{Graph-walk context compression.  BFS from question-anchored seed entities identifies structurally connected entities, assembled into compressed context organized by hop level.  No LLM calls required.}
\label{fig:gw}
\end{figure}

\paragraph{Phase 1: Seed entity identification.}
Question entities are anchored in the retrieved context via a multi-heuristic string-matching pipeline: (a)~exact substring matching (case-insensitive), (b)~multi-word matching where all content words of a multi-word entity appear in the question, and (c)~partial matching where a significant question word ($\geq$5 characters) appears as part of an entity name; common stop words are excluded.

\paragraph{Phase 2: Topological traversal.}
A breadth-first search (BFS) is performed from the seed entities through the knowledge graph, up to a depth of 3 hops.
Each reached entity is tagged with its hop distance from the nearest seed, establishing a structural proximity ranking.

\paragraph{Phase 3: Co-occurrence expansion.}
The BFS-reached entity set is expanded via a semantic bridge: if a text chunk mentions both a BFS-reached entity and an entity \emph{not} reached by BFS, the latter is added to the context.
This recovers entities that are semantically related but not directly connected in the extracted graph.

\paragraph{Phase 4: Priority-based assembly.}
The compressed context is assembled within a token budget $B$ (default $B{=}4{,}000$ tokens), prioritized by structural proximity to the question:
\begin{enumerate}
    \item Entity and relationship descriptions, ordered by hop distance from seeds (closer = higher priority)
    \item \emph{Tier-1 text chunks}: chunks mentioning BFS-reached entities, ranked by mention count
    \item \emph{Tier-2 text chunks}: chunks sharing question keywords but no BFS-reached entities
    \item Community report summaries, if present (lowest priority)
\end{enumerate}

Crucially, the output is reorganized by hop level, creating a natural interface with SPARQL CoT (\S\ref{sec:sparql_cot}): the hop-level sections align with the triple-pattern chain in the generated query, letting the model resolve each variable binding within a localized context region rather than scanning the entire input.
If no seed entities are matched, the algorithm falls back to the original uncompressed context.

The entire procedure is pure graph traversal and string matching, requiring no LLM calls or embedding computations.
In practice, the BFS reaches 14--21 entities on average, and the priority-based assembly achieves ${\sim}$60\% compression (${\sim}$10{,}000 $\to$ ${\sim}$4{,}000 tokens).
Figure~\ref{fig:gw} illustrates the four phases.

\section{Experimental Setup}
\label{sec:setup}

\subsection{Datasets}

We use three standard multi-hop QA benchmarks:%
\begin{itemize}
    \item \textbf{HotpotQA}~\cite{hotpotqa2018}: Bridge and comparison questions over Wikipedia paragraphs.
    Each question comes with 10 paragraphs (2 gold supporting paragraphs + 8 curated distractors).
    Bridge questions require following a chain of entities (e.g., ``The director of film X was born in what city?''), while comparison questions require attribute comparison (e.g., ``Which film was released first, X or Y?'').

    \item \textbf{MuSiQue}~\cite{musique2022}: Multi-hop questions requiring 2--4 reasoning steps, constructed by composing single-hop questions into chains.
    The deeper reasoning chains (up to 4 hops) are the primary source of difficulty: each additional hop compounds the chance of a reasoning failure.
    Each question comes with 20 paragraphs (2--4 gold supporting paragraphs + 16--18 distractors drawn from gold paragraphs of related questions, making them harder to distinguish than random distractors).
    Generally considered the hardest of the three benchmarks.

    \item \textbf{2WikiMultiHopQA}~\cite{2wikimhqa2020}: Questions constructed from Wikidata, requiring inference across two Wikipedia articles with the widest range of reasoning types among the three benchmarks (comparison, inference, compositional, and bridge).
    Each question comes with 10 paragraphs (2 or 4 gold supporting paragraphs + 6--8 distractors), following HotpotQA's format.
    KET-RAG's own evaluation does not include this dataset; we add it because of its type diversity.
\end{itemize}

For each dataset, we sample $N{=}500$ questions using a fixed seed (42) and pool all per-question paragraphs to construct the retrieval corpus.

\subsection{Model Stack}

Table~\ref{tab:models} shows our model configurations.
All LLM calls use the Groq inference platform, which provides fast inference for open-weight models at low per-token cost.

\begin{table}[t]
\centering
\small
\caption{Model stack. The index is built once with the 8B model and reused across all QA configurations.}
\label{tab:models}
\begin{tabular}{lll}
\toprule
\textbf{Component} & \textbf{Model} & \textbf{Cost} \\
\midrule
Embeddings & MiniLM-L6-v2 (384d) & Free \\
Index LLM & Llama-3.1-8B (Groq) & \$0.05/M \\
\midrule
QA (budget) & Llama-3.1-8B (Groq) & \$0.05/M \\
QA (standard) & Llama-3.3-70B (Groq) & \$0.59/M \\
\bottomrule
\end{tabular}
\end{table}

The index is model-independent: the knowledge graph and keyword indices do not depend on the QA model, allowing us to build once and vary the QA model without re-indexing.
Compared to KET-RAG's default stack (GPT-4o-mini + OpenAI embeddings), ours uses a $3\times$ cheaper LLM and free embeddings with $4\times$ lower dimensionality, for a total indexing cost of ${\sim}$\$1 per dataset.

At inference time, the 8B QA model is ${\sim}12\times$ cheaper per token than the 70B (\$0.05 vs.\ \$0.59/M tokens), motivating the question of whether augmentations can close the accuracy gap.

\subsection{Configurations}

We evaluate four QA configurations combining two prompting strategies with optional graph-walk compression:

\begin{enumerate}
    \item \textbf{Baseline:} The QA model receives the full KET-RAG-retrieved context (${\sim}$10{,}000 tokens) and answers directly.
    \item \textbf{Baseline + GW:} Same as Baseline but with graph-walk compressed context.
    \item \textbf{SPARQL CoT:} The QA model receives the full context with the SPARQL chain-of-thought prompt (\S\ref{sec:sparql_cot}).
    \item \textbf{SPARQL CoT + GW:} SPARQL CoT prompting applied to graph-walk compressed context.
\end{enumerate}

Each configuration is run with both the 8B and 70B QA models, yielding 8 runs per dataset and 24 runs total across the three benchmarks, for a total of 12{,}000 question-answer evaluations.
All QA calls use temperature 0.3.

\subsection{Evaluation}

We report accuracy, token-level F1, and exact match (EM) with SQuAD normalization.
Accuracy is determined by a heuristic-first pipeline: normalized string matching (lowercasing, stripping articles/punctuation, substring and last-token comparison), with LLM-judged semantic equivalence (Llama-3.1-8B) as fallback for remaining mismatches (e.g., ``NYC'' vs.\ ``New York City'').
A 102-judgment audit with Claude Opus~4.6 found 96.1\% agreement with our pipeline.
We also report abstain rate and context coverage (whether the gold answer appears in the retrieved context, separating retrieval from reasoning failures).
All comparisons are \emph{paired} on the same 500 questions and retrieved context for each dataset.


\section{Results}
\label{sec:results}

\subsection{Retrieval Quality: Coverage}
\label{sec:coverage}

With our budget components (free 384-dimensional embeddings, \$0.05/M-token indexing LLM), KET-RAG matches or exceeds its published coverage across all three benchmarks (Table~\ref{tab:coverage}), despite $3\times$ cheaper indexing LLM and $4\times$ lower embedding dimensionality.

\begin{table}[t]
\centering
\caption{Context coverage (\%) with budget components compared to KET-RAG~\cite{ketrag2025} with GPT-4o-mini + OpenAI embeddings.}
\label{tab:coverage}
\small
\setlength{\tabcolsep}{4pt}
\begin{tabular}{lccc}
\toprule
\textbf{Dataset} & \textbf{Ours} & \textbf{KET Low} & \textbf{KET High} \\
\midrule
HotpotQA & 90.8 (454/500) & 60.2 & 82.6 \\
MuSiQue & 77.2 (386/500) & 77.0 & 79.6 \\
2WikiMHQA & 81.0 (405/500) & --- & --- \\
\bottomrule
\end{tabular}
\end{table}

\subsection{QA Performance}
\label{sec:main_results}

Table~\ref{tab:main} presents accuracy, F1, and EM across all 24 configurations.

\begin{table*}[t]
\centering
\small
\setlength{\tabcolsep}{2.5pt}
\caption{Results (\%) on 500 paired questions per dataset. ``Covered'' = gold answer present in context (count in parentheses). \textbf{Bold} = best per column. GW = graph-walk compression. F1/EM use SQuAD normalization; 8B outputs are LLM-normalized to short answers (\S\ref{sec:setup}).}
\label{tab:main}
\begin{tabular}{ll|ccc|ccc|ccc||ccc|ccc|ccc}
\toprule
& & \multicolumn{9}{c||}{\textbf{All Questions}} & \multicolumn{9}{c}{\textbf{Covered Questions}} \\
& & \multicolumn{3}{c|}{Hotpot} & \multicolumn{3}{c|}{MuSiQue} & \multicolumn{3}{c||}{2Wiki} & \multicolumn{3}{c|}{Hotpot\,(454)} & \multicolumn{3}{c|}{MuSiQue\,(386)} & \multicolumn{3}{c}{2Wiki\,(405)} \\
\textbf{Method} & \textbf{Model} & Acc & F1 & EM & Acc & F1 & EM & Acc & F1 & EM & Acc & F1 & EM & Acc & F1 & EM & Acc & F1 & EM \\
\midrule
Base & 8B & 67.0 & 59.5 & 49.4 & 23.6 & 18.0 & 12.4 & 31.4 & 29.1 & 24.2 & 69.6 & 62.2 & 51.5 & 27.5 & 21.2 & 15.8 & 32.1 & 30.1 & 24.7 \\
~+GW & 8B & 63.6 & 55.9 & 46.0 & 19.4 & 14.6 & 9.0 & 30.4 & 28.1 & 24.8 & 66.3 & 58.3 & 48.0 & 22.6 & 17.0 & 11.4 & 33.1 & 30.8 & 26.9 \\
SPARQL & 8B & 70.6 & 60.8 & 49.4 & 28.8 & 22.6 & 15.4 & 45.6 & 36.9 & 31.0 & 74.4 & 64.2 & 52.6 & 34.8 & \textbf{27.4} & 19.7 & 49.9 & 40.6 & 33.6 \\
~+GW & 8B & \textbf{76.6} & \textbf{62.2} & \textbf{50.6} & \textbf{30.6} & \textbf{22.9} & \textbf{15.8} & \textbf{55.8} & \textbf{41.1} & \textbf{34.8} & \textbf{79.5} & \textbf{64.9} & \textbf{52.9} & \textbf{35.1} & 26.4 & \textbf{20.0} & \textbf{59.5} & \textbf{44.5} & \textbf{37.5} \\
\cmidrule{1-20}
Base & 70B & 78.0 & 67.9 & 56.6 & 35.2 & 26.9 & 18.6 & 48.8 & 43.0 & 36.8 & 81.3 & 70.9 & 59.0 & 40.3 & 32.5 & 23.9 & 54.8 & 47.9 & 40.7 \\
~+GW & 70B & 79.2 & 68.6 & 56.0 & 39.8 & 30.8 & 20.6 & 53.6 & 49.0 & 44.0 & 83.3 & 71.9 & 58.6 & 46.2 & 37.3 & 26.2 & 61.0 & 56.1 & 50.4 \\
SPARQL & 70B & \textbf{80.2} & \textbf{70.6} & \textbf{57.8} & \textbf{43.8} & 32.8 & 23.4 & \textbf{61.0} & \textbf{54.9} & \textbf{48.6} & \textbf{83.9} & \textbf{74.3} & \textbf{61.2} & \textbf{53.2} & 40.4 & 29.6 & \textbf{66.9} & \textbf{60.0} & \textbf{52.6} \\
~+GW & 70B & 79.6 & 70.5 & \textbf{57.8} & 42.2 & \textbf{34.0} & \textbf{25.8} & 59.8 & 51.5 & 44.0 & 83.5 & 74.0 & 60.8 & 51.4 & \textbf{42.3} & \textbf{32.5} & 66.7 & 57.6 & 49.6 \\
\bottomrule
\end{tabular}
\end{table*}

The headline result is that a budget 8B model, augmented with SPARQL CoT and graph-walk compression, surpasses the unaugmented 70B baseline on 2WikiMHQA (55.8\% vs.\ 48.8\%) and closes the gap on MuSiQue and HotpotQA, at ${\sim}12\times$ lower per-token cost.
SPARQL CoT alone improves \emph{every} configuration, with gains up to +14.2~pp on 2WikiMHQA for 8B and +12.2~pp for 70B.
When both augmentations are applied, 70B + SPARQL reaches the overall best accuracy (61.0\% on 2WikiMHQA, 80.2\% on HotpotQA).
Difficulty tracks reasoning depth as expected: HotpotQA (2-hop, 67-80\%), 2WikiMHQA (mixed, 30-61\%), MuSiQue (2-4 hop, 19-44\%).

The right half of the table conditions on covered questions (gold answer present in context), separating reasoning failures from retrieval gaps; we discuss this in \S\ref{sec:error_decomp}.
We analyze each factor below.

\subsection{Effect of SPARQL CoT Prompting}
\label{sec:sparql_effect}

Table~\ref{tab:sparql_delta} isolates the effect of SPARQL CoT by comparing it to the corresponding baseline (without graph-walk compression).

\begin{table}[t]
\centering
\small
\caption{Effect of SPARQL CoT: accuracy $\Delta$ (pp) and abstain $\Delta$ (pp) vs.\ baseline, without GW.}
\label{tab:sparql_delta}
\begin{tabular}{l|ccc}
\toprule
\textbf{Model} & \textbf{Hotpot} & \textbf{MuSiQue} & \textbf{2Wiki} \\
\midrule
\multicolumn{4}{c}{\textit{Accuracy $\Delta$ (pp)}} \\
\midrule
8B & $+3.6$ & $+5.2$ & $+14.2$ \\
70B & $+2.2$ & $+8.6$ & $+12.2$ \\
\midrule
\multicolumn{4}{c}{\textit{Abstain $\Delta$ (pp)}} \\
\midrule
8B & $-6.4$ & $-31.4$ & $-24.4$ \\
70B & $+0.4$ & $-8.2$ & $-4.2$ \\
\bottomrule
\end{tabular}
\end{table}

The accuracy gains correlate strongly with abstain rate reduction: on MuSiQue, 8B abstentions drop from 52.0\% to 20.6\% ({-}31.4~pp), suggesting that expressing relationships as triple patterns forces the model to \emph{engage} with multi-hop chains rather than declining, converting open-ended reasoning into structured pattern matching.

\subsection{CoT Ablation: SPARQL vs.\ Generic Decomposition}
\label{sec:cot_ablation}

We ablate whether the benefit stems from the SPARQL formalism or decomposition in general, comparing against a generic CoT prompt with natural-language sub-questions \rev{on all three datasets} (Table~\ref{tab:cot_ablation}).

\begin{table}[t]
\centering
\small
\caption{\rev{CoT ablation across all three datasets (without GW). Routing applies the 3-way classifier (\ref{app:routing_prompt}).}}
\label{tab:cot_ablation}
\setlength{\tabcolsep}{4pt}
\begin{tabular}{lcccc}
\toprule
\textbf{Method} & \textbf{8B Acc} & \textbf{8B Abs} & \textbf{70B Acc} & \textbf{70B Abs} \\
\midrule
\multicolumn{5}{c}{\rev{\textit{HotpotQA}}} \\
\rev{Baseline}    & \rev{67.0}          & \rev{12.2}          & \rev{78.0}          & \rev{7.8} \\
\rev{Generic CoT} & \rev{\textbf{75.6}} & \rev{7.8}           & \rev{79.8}          & \rev{11.2} \\
\rev{SPARQL CoT}  & \rev{70.6}          & \rev{5.8}           & \rev{80.2}          & \rev{8.2} \\
\rev{Routing}     & \rev{75.0}          & \rev{\textbf{1.2}}  & \rev{\textbf{82.4}} & \rev{\textbf{6.6}} \\
\midrule
\multicolumn{5}{c}{\textit{2WikiMHQA}} \\
Baseline    & 31.4 & 46.0 & 48.8 & 34.8 \\
Generic CoT & 52.6 & 25.2 & 56.4 & 36.8 \\
SPARQL CoT  & 45.6 & 21.6 & 61.0 & 30.6 \\
Routing     & \textbf{58.4} & \textbf{14.2} & \textbf{66.4} & \textbf{26.8} \\
\midrule
\multicolumn{5}{c}{\rev{\textit{MuSiQue}}} \\
\rev{Baseline}    & \rev{23.6}          & \rev{52.0}           & \rev{35.2}          & \rev{42.4} \\
\rev{Generic CoT} & \rev{\textbf{37.0}} & \rev{29.2}           & \rev{38.2}          & \rev{41.4} \\
\rev{SPARQL CoT}  & \rev{28.8}          & \rev{20.6}           & \rev{43.8}          & \rev{34.2} \\
\rev{Routing}     & \rev{33.8}          & \rev{\textbf{10.2}}  & \rev{\textbf{47.0}} & \rev{\textbf{28.6}} \\
\bottomrule
\end{tabular}
\end{table}

\rev{Three patterns hold across all three datasets.}

\rev{First, both CoT methods improve substantially over the baseline on every dataset, confirming that structured decomposition is the primary driver of the gains.}

\rev{Second, question-type routing is the best of the four methods on 70B across all three datasets (HotpotQA 82.4, 2WikiMHQA 66.4, MuSiQue 47.0). A per-native-type breakdown (Appendix~\ref{app:per_type_sparql_vs_generic}) shows why: bridge and structural types, which constitute the majority of questions on each benchmark, favour SPARQL on 70B, while comparison and inference questions favour generic CoT on the datasets where they exist. Routing exploits this complementarity.}

\rev{Third, the SPARQL-vs-generic ranking depends on model size. 70B reliably benefits from SPARQL on all three datasets (80.2 / 61.0 / 43.8 vs.\ 79.8 / 56.4 / 38.2), while 8B without GW prefers generic CoT on HotpotQA and MuSiQue and is closer on 2WikiMHQA. A hint at the cause comes from a syntax-overhead mechanism: 8B's SPARQL parse rate averages 87.4\% vs.\ 93.4\% for 70B, with the largest per-dataset gap on HotpotQA (10.8~pp), exactly where 8B's SPARQL/generic accuracy gap is also largest (Appendix~\ref{app:syntax_validity}).}

\rev{On 8B without GW, routing wins on 2WikiMHQA, ties generic on HotpotQA, and trails on MuSiQue. With GW added (Table~\ref{tab:routing}), routing wins on all datasets, with the largest gain of +16.0~pp on 2WikiMHQA.}

Notably, both CoT methods improve accuracy for the 8B model \rev{on every dataset}, contrasting with findings that CoT primarily helps larger models~\cite{wei2022cot}.

\rev{\paragraph{Comparison with Self-Ask.} Among three established multi-hop CoT methods (IRCoT~\cite{trivedi2023ircot}, Self-Ask~\cite{press2023selfask}, Decomposed Prompting~\cite{khot2023decomposed}), only Self-Ask has a published single-call form; single-call adaptations of the other two collapse to our generic CoT and to Self-Ask respectively, since their distinguishing components (interleaved retrieval; sub-task delegation) are multi-call by construction. Under apples-to-apples prompting (Press et al.'s Bamboogle precedent of non-test-tuned prompts on both methods), SPARQL CoT wins 5 of 6 dataset-model cells by $+6.7$~pp on average for those cells; the single loss is $-$3.0~pp on MuSiQue 70B (Appendix~\ref{app:self_ask}).}

\subsection{Question-Type Routing}
\label{sec:routing}

Because 8B's preferred CoT variant depends on question type (\S\ref{sec:cot_ablation}), we deploy the routing policy across all three datasets with graph-walk compression (Table~\ref{tab:routing}).

\begin{table}[t]
\centering
\small
\setlength{\tabcolsep}{2.5pt}
\caption{8B with routing + GW vs.\ unaugmented 70B baseline on KET-RAG. Routing classifies each question, applies the preferred CoT, and falls back on abstain.}
\label{tab:routing}
\begin{tabular}{l|ccc|ccc}
\toprule
& \multicolumn{3}{c|}{\textbf{Accuracy (\%)}} & \multicolumn{3}{c}{\textbf{Abstain (\%)}} \\
\textbf{Dataset} & \textbf{8B} & \textbf{70B} & \textbf{$\Delta$} & \textbf{8B} & \textbf{70B} & \textbf{$\Delta$} \\
\midrule
HotpotQA & \textbf{79.8} & 78.0 & $+1.8$ & \textbf{1.0} & 5.6 & $-4.6$ \\
MuSiQue & \textbf{35.6} & 35.2 & $+0.4$ & \textbf{6.2} & 16.2 & $-10.0$ \\
2WikiMHQA & \textbf{64.8} & 48.8 & $\mathbf{+16.0}$ & \textbf{7.2} & 34.8 & $\mathbf{-27.6}$ \\
\bottomrule
\end{tabular}
\end{table}

With routing and graph-walk compression, the 8B model matches or exceeds the unaugmented 70B baseline on all three benchmarks at ${\sim}12\times$ lower cost.
The gains are largest on 2WikiMHQA (+16.0~pp), where question-type diversity (bridge, comparison, inference, compositional) gives routing the most leverage.
Without routing, SPARQL+GW alone surpasses 70B only on 2WikiMHQA (+7.0~pp) while trailing on HotpotQA ({-}1.4) and MuSiQue ({-}4.6); routing closes these remaining gaps.
The routing policy adds only a cheap classifier call (a few tokens) and at most one retry per question, keeping total inference cost well below a single 70B call.

\paragraph{Implementation.}
The router is a single 8B call (max 5 output tokens) that classifies each question as \emph{bridge}, \emph{comparison}, or \emph{inference} (exact prompt in Appendix~\ref{app:routing_prompt}).
Bridge questions go to SPARQL CoT; comparison and inference questions go to generic CoT.
If the chosen method abstains, the question is retried with the other method.
The full pipeline thus requires at most three 8B calls per question: classify, answer, and optionally retry.

\subsection{Effect of Graph-Walk Compression}
\label{sec:gw_effect}

Graph-walk compression shows a more complex interaction with model capability and prompting strategy (Table~\ref{tab:gw_delta}).

\begin{table}[t]
\centering
\small
\setlength{\tabcolsep}{4pt}
\caption{Effect of graph-walk compression: accuracy $\Delta$ (pp) vs.\ non-GW config.}
\label{tab:gw_delta}
\begin{tabular}{ll|ccc|c}
\toprule
\textbf{Method} & \textbf{Model} & \textbf{Hot.} & \textbf{Mus.} & \textbf{2Wi.} & \textbf{Avg} \\
\midrule
Base & 8B & $-3.4$ & $-4.2$ & $-1.0$ & $-2.9$ \\
Base & 70B & $+1.2$ & $+4.6$ & $+4.8$ & $+3.5$ \\
SPARQL & 8B & $+6.0$ & $+1.8$ & $+10.2$ & $+6.0$ \\
SPARQL & 70B & $-0.6$ & $-1.6$ & $-1.2$ & $-1.1$ \\
\bottomrule
\end{tabular}
\end{table}

A consistent pattern emerges: GW helps most for SPARQL + 8B (+6.0~pp avg, +10.2 on 2WikiMHQA) and Base + 70B (+3.5~pp avg), but hurts Base + 8B ({-}2.9~pp).
For SPARQL + 70B, GW incurs only a {-}1.1~pp average loss while reducing input tokens by ${\sim}$60\% (${\sim}$10{,}000 $\to$ ${\sim}$4{,}000), a favorable accuracy-cost trade-off given the 70B model's higher per-token price.
This reveals a complementarity: GW helps when the model has structured guidance (SPARQL + 8B) or sufficient capacity to reason over reduced noise (70B baseline); without either, the small model relies on exhaustive context for pattern matching and compression removes useful signal.

\rev{Against truncation and top-k embedding-similarity baselines at the same 4k-token budget on 8B, GW outperforms both on SPARQL CoT across all three datasets ($+9.2$ / $+8.8$ / $+3.2$~pp over top-k), supporting GW's structural alignment with triple-pattern reasoning (Appendix~\ref{app:compression_baselines}).}

\subsection{Error Decomposition}
\label{sec:error_decomp}

To disentangle retrieval and reasoning failures, we decompose errors by whether the gold answer was present in the retrieved context.
Table~\ref{tab:error_decomp} shows the reasoning share of total errors. The ``Covered'' columns of Table~\ref{tab:main} report accuracy, F1, and EM conditioned on coverage across all configurations.

\begin{table}[t]
\centering
\small
\caption{Reasoning share of total errors (\%): incorrect answers where the gold answer \emph{was} present in context.}
\label{tab:error_decomp}
\begin{tabular}{lccc}
\toprule
\textbf{Config} & \textbf{Hotpot} & \textbf{MuSiQue} & \textbf{2Wiki} \\
\midrule
8B Baseline & 84 & 73 & 80 \\
70B + SPARQL CoT & 74 & 64 & 69 \\
\bottomrule
\end{tabular}
\end{table}

Reasoning failures dominate: on MuSiQue for example, the answer is present in context for 77.2\% of questions (Table~\ref{tab:coverage}), yet the 8B baseline answers only 23.6\% correctly overall, a gap of over 50 points attributable to reasoning failures.
SPARQL CoT with 70B closes roughly half this gap (53.2\% on covered questions), confirming that the dominant bottleneck in Graph-RAG QA is reasoning, not retrieval.

On HotpotQA, 70B + SPARQL CoT reaches 83.9\% on covered questions, approaching the 90.8\% coverage ceiling, suggesting that for 2-hop questions the remaining errors are mostly retrieval gaps. \rev{For error decomposition specifically, a stricter bridging-entity coverage metric, requiring all entities needed to derive the answer in the retrieved context (Appendix~\ref{app:m3_coverage}), gives 88.2 / 66.0 / 69.4\% coverage and confirms reasoning failures remain the dominant source in 5 of 6 dataset-model cells.}

\subsection{Analysis by Question Type}
\label{sec:per_type}

\rev{A per-type breakdown (Appendix~\ref{app:per_type_gains}) confirms that SPARQL CoT's gains concentrate on chain-following types (bridge\_comparison sees $+18.2$ / $+28.1$~pp on 2WikiMultiHopQA for 8B/70B) and are smallest on inference ($+3.7$~pp), consistent with the SPARQL formalism's structural match. GW compression similarly benefits the deepest hops (+10 to +12~pp on MuSiQue 3-hop and 4-hop) and inference questions ($+16.7$~pp on 2WikiMultiHopQA), where focused context most helps integration across entities.}

\section{Discussion}
\label{sec:discussion}

\paragraph{Reasoning, not retrieval, is the bottleneck.}
Interestingly, KET-RAG achieves 77\% to 91\% coverage even with budget components (\S\ref{sec:coverage}), yet accuracy lags far behind: the coverage-to-accuracy gap exceeds 50 percentage points on MuSiQue (\S\ref{sec:error_decomp}).
As the graph index is model-independent, we can build once with inexpensive models and upgrade the QA model without re-indexing.

\paragraph{Structured decomposition is the primary lever.}
Our key findings show that SPARQL CoT improves every configuration we tested (\S\ref{sec:sparql_cot}), including 8B models, contrasting with findings that chain-of-thought primarily benefits models above ${\sim}$100B parameters~\cite{wei2022cot}.
The ablation (\S\ref{sec:cot_ablation}) shows the benefit comes from decomposition itself; the SPARQL formalism adds further value for 70B.

\paragraph{How far can a small model go?}
8B + SPARQL CoT + GW surpasses the unaugmented 70B baseline on 2WikiMHQA (55.8\% vs.\ 48.8\%) at ${\sim}12\times$ lower per-token cost.
On MuSiQue, however, 70B retains a +4.6~pp advantage, suggesting that deeper reasoning chains (3--4 hops) require larger models, while 2-hop questions can be scaffolded effectively even with small models.
Question-type routing (\S\ref{sec:routing}) pushes this further: by steering each question to the appropriate CoT variant and retrying on abstention, the fully augmented 8B surpasses unaugmented 70B on all three benchmarks.

\paragraph{Generality beyond KET-RAG.}
To test whether these results are system-specific, we replicate the experiment on LightRAG~\cite{lightrag2024} across all three benchmarks (Appendix~\ref{app:lightrag}).
SPARQL CoT consistently improves LightRAG: +7.6~pp on 2WikiMHQA, +5.4~pp on MuSiQue, and +2.2~pp on HotpotQA, confirming that structured reasoning transfers whenever the context contains entity-relationship structure.
\rev{Graph-walk compression, by contrast, hurts on LightRAG, consistent with its hybrid retrieval lacking the $\beta$-stratified relevance gradient that GW traversal exploits in KET-RAG.}
This suggests SPARQL CoT is \rev{a Graph-RAG-family-agnostic prompting strategy whenever the retrieved context exposes triple-structured entity-relationship content}, while GW compression requires a retrieval-pipeline relevance gradient to be effective. \rev{Retrievers that return flat passages or summarized community reports (e.g., HippoRAG, KGP) fall outside this structural condition and would require a different formalism.}

\paragraph{Deployment recommendations.}
The results suggest a simple decision rule.
For budget deployments (8B), always enable routing with graph-walk compression: it matches or exceeds 70B at ${\sim}12\times$ lower cost, with negligible overhead (one classifier call plus at most one fallback).
For quality-maximizing deployments (70B), use SPARQL CoT without compression; GW's ${\sim}$1~pp accuracy cost rarely justifies the token savings at this scale.
GW should be avoided only for Base + 8B, where the small model lacks reasoning guidance to navigate compressed context.

\section{Related Work}
\label{sec:related}

\paragraph{Graph-RAG systems.}
\rev{A key distinction across Graph-RAG systems is whether the retrieved context preserves entity-relationship (E-R) structure.}
GraphRAG~\cite{graphrag2024}, KET-RAG~\cite{ketrag2025}, and LightRAG~\cite{lightrag2024} return entity descriptions, relationship triples, and text chunks, making their context amenable to structured prompting and graph-based compression.
HybridRAG~\cite{hybridrag2024} also combines KG subgraphs with vector-retrieved passages, though it has only been evaluated on financial transcripts.
Other systems use graph structure for retrieval but return flat passages: HippoRAG~\cite{hipporag2024,hipporag2_2025}, KGP~\cite{knngrag2024}, Clue-RAG~\cite{cluerag2025}, and GraphRAG-V~\cite{yu2025graphragv}.
\citet{whentousegraphs2025} identify context inflation as a key failure mode.
All the above target retrieval quality; we study reasoning over the retrieved E-R context with budget models.

\paragraph{Chain-of-thought prompting.}
\rev{IRCoT~\cite{trivedi2023ircot} and Decomposed Prompting~\cite{khot2023decomposed} are multi-call by construction; their single-call adaptations collapse to our generic CoT and to Self-Ask~\cite{press2023selfask} respectively, which we compare against directly (\S\ref{sec:cot_ablation}, Appendix~\ref{app:self_ask}).}
SPARQL has been used for knowledge base QA~\cite{gu2023dont} but never as a CoT scaffold for Graph-RAG.

\paragraph{Context compression.}
LLMLingua~\cite{llmlingua2023} and RECOMP~\cite{xu2024recomp} compress passages via learned models, incurring inference cost.
Graph-RAG context has exploitable topology that enables LLM-free compression via BFS.
The interaction between compression and structured prompting has not been previously studied.

\section{Conclusion}
\label{sec:conclusion}

Once retrieval achieves high coverage, the dominant bottleneck in Graph-RAG QA is reasoning.
We propose three inference-time augmentations that target this gap: SPARQL CoT (+2 to +14~pp), graph-walk compression, and question-type routing; none requires retraining or re-indexing.
Together, they enable a fully augmented 8B model to match or exceed an unaugmented 70B baseline on all three benchmarks at ${\sim}12\times$ lower cost.
Because the graph index is model-independent, stronger future models can be swapped in without re-indexing.
For budget deployments, routing with graph-walk compression matches or exceeds 70B; for quality-maximizing deployments \rev{on 70B}, SPARQL CoT alone suffices.
Code (MIT): \url{https://github.com/thomouvic/graph-rag-qa-pub}

\newpage
\section*{Limitations}

We evaluate with two Llama-3 sizes and a single embedding model; \rev{cross-architecture experiments (e.g., Mistral, Qwen) would better characterize generality.}

\rev{In SPARQL CoT, the model writes the full reasoning chain in one LLM call. If an early step fails (e.g., the first variable cannot be bound to any entity in the context), the rest of the chain inherits that failure; we detect this via the abstain mechanism, but do not retry. Multi-call recovery falls outside the paper's single-call cost-efficient scope and is future work.}

\rev{Our SPARQL parse-rate analysis (Appendix~\ref{app:syntax_validity}) measures structural-syntax validity only; we have no direct predicate-grounding metric.}

We use the default $\theta = 0.5$ for KET-RAG's entity-keyword balance, \rev{where per-dataset tuning might improve results. A $\beta{=}0$ ablation of retrieval stratification and a soft NL-infused structured-prompt variant are natural next steps.}

We used an AI coding assistant (Claude Code) to help with experiment scripting and paper editing; all scientific contributions were made by the authors.

\bibliography{custom}

\newpage
\appendix

\section{Appendix}

\begin{table*}[t!]
\centering
\footnotesize
\setlength{\tabcolsep}{2.2pt}
\caption{LightRAG generality experiment (\%) on 500 paired questions per dataset. ``Covered'' = gold answer present in LightRAG context (count in parentheses). \textbf{Bold} = best per model group. GW = graph-walk compression. F1/EM use SQuAD normalization; 8B outputs are LLM-normalized to short answers (\S\ref{sec:setup}).}
\label{tab:lightrag}
\begin{tabular}{ll|ccc|ccc|ccc||ccc|ccc|ccc}
\toprule
& & \multicolumn{9}{c||}{\textbf{All Questions}} & \multicolumn{9}{c}{\textbf{Covered Questions}} \\
& & \multicolumn{3}{c|}{Hotpot} & \multicolumn{3}{c|}{MuSiQue} & \multicolumn{3}{c||}{2Wiki} & \multicolumn{3}{c|}{Hotpot\,(450)} & \multicolumn{3}{c|}{MuSiQue\,(374)} & \multicolumn{3}{c}{2Wiki\,(417)} \\
\textbf{Method} & \textbf{Model} & Acc & F1 & EM & Acc & F1 & EM & Acc & F1 & EM & Acc & F1 & EM & Acc & F1 & EM & Acc & F1 & EM \\
\midrule
Base & 8B & 62.2 & 55.1 & 45.2 & 26.6 & 19.3 & 12.8 & 32.2 & 29.0 & 24.6 & 65.8 & 57.8 & 47.3 & 34.2 & 23.7 & 16.6 & 34.5 & 31.2 & 26.6 \\
~+GW & 8B & 54.4 & 47.8 & 38.4 & 25.4 & 18.1 & 11.8 & 35.2 & 31.9 & 27.6 & 56.0 & 49.1 & 39.1 & 30.7 & 22.0 & 15.2 & 35.7 & 32.2 & 27.6 \\
SPARQL & 8B & \textbf{66.8} & \textbf{57.5} & \textbf{46.6} & \textbf{37.4} & \textbf{26.9} & \textbf{18.6} & \textbf{57.4} & 44.3 & 36.2 & \textbf{69.6} & \textbf{59.3} & \textbf{47.8} & \textbf{46.3} & \textbf{33.0} & \textbf{24.1} & \textbf{61.9} & \textbf{47.1} & 38.8 \\
~+GW & 8B & 66.2 & 55.2 & 44.4 & 33.6 & 23.5 & 15.8 & 56.4 & \textbf{44.6} & \textbf{37.8} & 69.3 & 57.0 & 45.6 & 40.1 & 28.1 & 20.3 & 60.2 & 46.6 & \textbf{39.8} \\
\cmidrule{1-20}
Base & 70B & 79.8 & 68.1 & 55.6 & 41.6 & 32.0 & 24.2 & 59.8 & 49.1 & 41.8 & 83.3 & 70.7 & 57.8 & 52.1 & 40.2 & 31.8 & 64.5 & 52.1 & 44.8 \\
~+GW & 70B & 68.8 & 59.9 & 49.4 & 35.0 & 25.7 & 17.2 & 51.2 & 45.8 & 39.2 & 71.8 & 62.2 & 51.1 & 44.4 & 32.5 & 23.0 & 54.9 & 49.0 & 42.0 \\
SPARQL & 70B & \textbf{82.0} & \textbf{71.6} & \textbf{57.8} & \textbf{47.0} & \textbf{36.5} & \textbf{28.0} & \textbf{67.4} & \textbf{58.1} & \textbf{49.0} & \textbf{85.8} & \textbf{74.2} & \textbf{59.8} & \textbf{58.8} & \textbf{46.6} & \textbf{36.9} & \textbf{72.4} & \textbf{62.7} & \textbf{52.8} \\
~+GW & 70B & 64.8 & 56.5 & 45.4 & 36.6 & 29.7 & 23.6 & 50.2 & 45.0 & 37.6 & 67.3 & 58.2 & 46.4 & 47.1 & 38.4 & 31.3 & 53.7 & 47.8 & 40.0 \\
\bottomrule
\end{tabular}
\end{table*}

\noindent
This appendix provides supplementary analysis and experimental details referenced in the main paper.
\S\ref{app:lightrag} presents the full LightRAG generality experiment summarized in \S\ref{sec:discussion}.
\S\ref{app:routing_prompt} gives the routing classifier prompt used in \S\ref{sec:routing}.
\rev{\S\ref{app:generic_cot_prompt} gives the generic CoT prompt used as a baseline throughout the paper.}
\rev{\S\ref{app:self_ask} reports the full Self-Ask comparison summarized in \S\ref{sec:cot_ablation}, with adaptation notes on IRCoT and Decomposed Prompting.}
\rev{\S\ref{app:m3_coverage} presents the bridging-entity coverage setup and per-cell results referenced in \S\ref{sec:error_decomp}.}
\rev{\S\ref{app:syntax_validity} reports per-cell SPARQL parse rates supporting the syntax-overhead discussion in \S\ref{sec:cot_ablation}.}
\rev{\S\ref{app:per_type_sparql_vs_generic} gives the per-native-type SPARQL CoT vs generic CoT breakdown referenced in \S\ref{sec:cot_ablation}.}
\rev{\S\ref{app:per_type_gains} presents the per-question-type SPARQL CoT and GW gain tables referenced in \S\ref{sec:per_type}.}
\rev{\S\ref{app:compression_baselines} compares graph-walk compression against simpler alternatives (truncation, top-k embedding similarity) at the same budget.}

\subsection{Generality: LightRAG Experiment}
\label{app:lightrag}

To test whether SPARQL CoT generalises beyond KET-RAG, we replicate the full experiment on LightRAG~\cite{lightrag2024}, a prominent open-source Graph-RAG system, across all three benchmarks.
We use the same models as our KET-RAG setup (Llama-3.1-8B for indexing, both 8B and 70B for QA) and retrieve hybrid (keyword + vector) context for each of the 500 questions, then run all eight QA configurations (4 methods $\times$ 2 models).
LightRAG context coverage is 450/500 on HotpotQA, 374/500 on MuSiQue, and 417/500 on 2WikiMHQA.

Table~\ref{tab:lightrag} mirrors Table~\ref{tab:main} for direct comparison.
SPARQL CoT consistently improves all configurations, with larger gains for 8B, confirming that structured decomposition disproportionately benefits smaller models.
The overall gains are smaller than on KET-RAG, likely because LightRAG aggregates entities into high-level descriptions during extraction, making triple-pattern decomposition a less precise match for the retrieved context.

Graph-walk compression, by contrast, \emph{hurts} on LightRAG across all configurations, the opposite of KET-RAG.
Because the two systems retrieve similar-sized contexts (median ${\sim}13{,}000$ vs.\ ${\sim}10{,}600$ tokens), the difference is not one of context length but of context \emph{structure}: KET-RAG's retrieval separates core chunks ($\beta{=}0.8$) from peripheral ones, creating a relevance gradient that graph-walk traversal can exploit.
LightRAG's hybrid retrieval, by contrast, returns material without such stratification, so pruning removes relevant content as readily as noise.

These results support our central claim that SPARQL CoT \rev{is agnostic across Graph-RAG systems that expose triple-structured entity-relationship content}: it improves multi-hop QA on both KET-RAG and LightRAG, regardless of which produced the context.
The GW finding adds a practical caveat: compression is most beneficial when the retrieval pipeline introduces a clear relevance gradient that graph-walk traversal can exploit, rather than being determined by context size alone.

\subsection{Routing Classifier Prompt}
\label{app:routing_prompt}

The question-type classifier (\S\ref{sec:routing}) uses the following prompt with the same 8B model (max 5 output tokens):

\begin{quote}
\small
\texttt{Classify this question as exactly one of: "bridge", "comparison", "inference".} \\
\texttt{Definitions:} \\
\texttt{- bridge: answer requires following an entity chain across facts} \\
\texttt{- comparison: answer compares two entities/values} \\
\texttt{- inference: answer requires implicit reasoning not a clean entity chain} \\
\texttt{Reply with exactly one word: bridge or comparison or inference} \\[4pt]
\texttt{Question: \{question\}}
\end{quote}

If the model returns a label outside \{bridge, comparison, inference\}, the system defaults to bridge.

\subsection{\rev{Generic CoT Prompt}}
\label{app:generic_cot_prompt}

\rev{The generic CoT baseline (\S\ref{sec:cot_ablation}) uses the following single-call decomposition prompt:}

\begin{quote}
\small
\texttt{You are answering a multi-hop question using ONLY the provided context.} \\[2pt]
\texttt{Step 1:} Break down the question into simpler sub-questions that need to be answered one at a time. \\
\texttt{Step 2:} For each sub-question, find the relevant information in the context and write the intermediate answer. \\
\texttt{Step 3:} Combine the intermediate answers to reach the final answer. \\
\texttt{Step 4:} Write \texttt{FINAL ANSWER: <your answer in a few words>} \\[2pt]
If the answer is not in the context, write \texttt{FINAL ANSWER: I don't know} \\[2pt]
\texttt{CONTEXT: \{context\}} \\
\texttt{QUESTION: \{question\}}
\end{quote}

\subsection{\rev{Self-Ask Comparison: Setup and Full Results}}
\label{app:self_ask}

\rev{We compare SPARQL CoT against Self-Ask~\cite{press2023selfask} using the verbatim few-shot prompts from Press et al.'s Tables~13 (4-shot) and~10 (6-shot).}

\rev{\paragraph{Adaptation for retrieval.} Press et al.\ evaluate Self-Ask with parametric knowledge or an external search engine. Our setting is retrieval-grounded, so we insert the retrieved text as a labelled \texttt{Context:} block between the few-shot demonstrations and the test question. This is the only adaptation. Following Press et al.'s footnote~4, we append \texttt{Are follow up questions needed here: Yes.} to the prompt for smaller-than-Davinci models (both Llama-3.1-8B and Llama-3.3-70B qualify).}

\rev{\paragraph{Decoding.} Temperature 0.3, max\_tokens 2048. Both methods receive the same retrieved context, without GW compression.}

\rev{\paragraph{Symmetric prompts (no test-set tuning).} For an apples-to-apples comparison, neither method's prompt is tuned to the test dataset. Self-Ask uses Press et al.'s 4-shot prompt across all three test datasets, following Press et al.'s own Bamboogle precedent of prompt transfer. SPARQL CoT uses its generic single-example prompt uniformly.}

\begin{table}[t]
\centering
\small
\setlength{\tabcolsep}{3pt}
\rev{\caption{Self-Ask vs.\ SPARQL CoT, full accuracies (\%) on the same 500 questions per dataset, both methods using non-test-tuned prompts.}}
\rev{\label{tab:self_ask}}
\begin{tabular}{llrrr}
\toprule
\rev{\textbf{Dataset}} & \rev{\textbf{Model}} & \rev{\textbf{S-Ask}} & \rev{\textbf{SPARQL}} & \rev{\textbf{Gap (pp)}} \\
\midrule
\rev{HotpotQA}   & \rev{8B}  & \rev{57.0} & \rev{\textbf{70.6}} & \rev{$+13.6$} \\
\rev{HotpotQA}   & \rev{70B} & \rev{70.0} & \rev{\textbf{80.2}} & \rev{$+10.2$} \\
\rev{2WikiMHQA}  & \rev{8B}  & \rev{42.2} & \rev{\textbf{45.6}} & \rev{$+3.4$}  \\
\rev{2WikiMHQA}  & \rev{70B} & \rev{58.6} & \rev{\textbf{61.0}} & \rev{$+2.4$}  \\
\rev{MuSiQue}    & \rev{8B}  & \rev{25.0} & \rev{\textbf{28.8}} & \rev{$+3.8$}  \\
\rev{MuSiQue}    & \rev{70B} & \rev{\textbf{46.8}} & \rev{43.8} & \rev{$-3.0$}  \\
\bottomrule
\end{tabular}
\end{table}

\rev{\paragraph{Adaptation notes (IRCoT, Decomposed Prompting).} Of three established multi-hop methods (IRCoT~\cite{trivedi2023ircot}, Self-Ask, Decomposed Prompting~\cite{khot2023decomposed}), only Self-Ask has a published single-call form. The other two are defined by their multi-call interactions: IRCoT by interleaved retrieval, Decomposed Prompting by sub-task handler delegation. A single-call adaptation of either strips out the contribution that distinguishes it from chain-of-thought: single-call IRCoT collapses to chain-of-thought over the up-front context (which we test via generic CoT), and single-call Decomposed Prompting collapses to Self-Ask's decompose-and-answer pattern (which we test directly). We therefore selected Self-Ask as the established single-call baseline.}

\subsection{\rev{Bridging-Entity Coverage: Setup and Results}}
\label{app:m3_coverage}

\rev{The substring coverage metric used in Tables~\ref{tab:coverage} and~\ref{tab:error_decomp} counts a question as covered if the gold answer string appears in the retrieved context. It measures retrieval quality directly and matches the metric used by KET-RAG and prior RAG work. It can, however, both overcount (the gold string appearing in a distractor) and undercount (the answer paraphrased).}

\rev{For error decomposition specifically, we additionally compute a stricter \emph{bridging-entity} coverage metric that requires all entities needed to derive the answer to appear in the retrieved context (case-insensitive substring). Each benchmark publishes its supporting-fact annotations in a different format, so the bridging-entity set is constructed using each benchmark's native annotation:}

\begin{itemize}
\item \rev{\textbf{HotpotQA:} the article titles of the two supporting paragraphs (the bridging entities in HotpotQA's annotation convention), plus the final answer.}
\item \rev{\textbf{2WikiMultiHopQA:} the subjects and objects of all evidence triples (i.e., the entities along the explicit reasoning chain).}
\item \rev{\textbf{MuSiQue:} all intermediate decomposition-step answers (the bridging values produced en route to the final answer).}
\end{itemize}

\rev{The average bridging-entity set size is 3.0 on HotpotQA, 4.0 on 2WikiMultiHopQA, and 2.7 on MuSiQue. Coverage under the bridging-entity metric is 88.2\% on HotpotQA, 66.0\% on 2WikiMultiHopQA, and 69.4\% on MuSiQue. The steeper drop on 2WikiMultiHopQA (from 81.0\% under substring matching) reflects its larger average set size, since covering 4 entities per question is strictly harder than covering 3 or 2.7.}

\begin{table}[t]
\centering
\small
\setlength{\tabcolsep}{4pt}
\rev{\caption{Reasoning share of total errors (\%) under the bridging-entity coverage metric, per dataset-model cell. Values $>$50\% indicate reasoning is the dominant error source.}}
\label{tab:m3_reasoning}
\begin{tabular}{llr}
\toprule
\rev{\textbf{Dataset}} & \rev{\textbf{Model}} & \rev{\textbf{Reasoning share (\%)}} \\
\midrule
\rev{HotpotQA}         & \rev{8B}  & \rev{78.8} \\
\rev{HotpotQA}         & \rev{70B} & \rev{70.9} \\
\rev{2WikiMultiHopQA}  & \rev{8B}  & \rev{58.0} \\
\rev{2WikiMultiHopQA}  & \rev{70B} & \rev{46.5} \\
\rev{MuSiQue}          & \rev{8B}  & \rev{63.6} \\
\rev{MuSiQue}          & \rev{70B} & \rev{60.2} \\
\bottomrule
\end{tabular}
\end{table}

\rev{Reasoning failures remain the dominant source of errors in 5 of 6 dataset-model cells. The only flip is 2WikiMultiHopQA 70B, where retrieval narrowly dominates (53.5\% retrieval vs.\ 46.5\% reasoning). The direction of the central claim is robust to the choice of metric.}

\subsection{\rev{SPARQL Syntax-Validity Rates}}
\label{app:syntax_validity}

\rev{To quantify the syntax-overhead mechanism noted in \S\ref{sec:cot_ablation}, we measure the parse rate of the SPARQL block produced by each SPARQL CoT response (Table~\ref{tab:syntax_validity}). From each model output we extract the first \texttt{SELECT ... \{ ... \}} block (brace-balanced) and check whether it parses under rdflib's SPARQL grammar. Because the model uses bare predicate names like \texttt{tributaryOf} (we deliberately give it no schema), we first wrap bare names in IRI brackets \texttt{<...>} before parsing. The metric therefore measures \emph{structural} grammar validity (well-formed SELECT, WHERE, and triple-pattern syntax), not predicate vocabulary correctness.}

\begin{table}[t]
\centering
\small
\setlength{\tabcolsep}{4pt}
\caption{\rev{SPARQL parse rate (\%) per dataset and model. Both models attempt a SPARQL query on essentially every question (extraction rate 96--99\%); the rate is computed over all 500 questions.}}
\label{tab:syntax_validity}
\begin{tabular}{lrr}
\toprule
\rev{\textbf{Dataset}} & \rev{\textbf{8B}} & \rev{\textbf{70B}} \\
\midrule
\rev{HotpotQA}         & \rev{81.8} & \rev{92.6} \\
\rev{2WikiMultiHopQA}  & \rev{91.4} & \rev{95.4} \\
\rev{MuSiQue}          & \rev{89.0} & \rev{92.2} \\
\midrule
\rev{Average}          & \rev{87.4} & \rev{93.4} \\
\bottomrule
\end{tabular}
\end{table}

\rev{Both models produce mostly valid SPARQL (averages 87.4\% on 8B, 93.4\% on 70B), but 8B is consistently more malformed, and the gap is largest on HotpotQA (10.8~pp), where 8B's SPARQL CoT vs.\ generic CoT accuracy deficit is also largest (\S\ref{sec:cot_ablation}). This supports the hypothesised syntax-overhead mechanism but does not establish it as the sole cause; alternative explanations (longer reasoning traces in generic CoT contributing via test-time compute, better alignment of natural-language CoT with smaller models' post-training distributions) remain plausible.}

\subsection{\rev{Per-Native-Type SPARQL CoT vs Generic CoT}}
\label{app:per_type_sparql_vs_generic}

\rev{This appendix gives the per-native-type breakdown of SPARQL CoT vs generic CoT (without GW) that underpins the routing rule (\S\ref{sec:routing}), shown in Table~\ref{tab:per_type_sparql_vs_generic}. The two methods are complementary rather than competing: SPARQL is strongest on structural types (where triple-pattern decomposition aligns with the entity-relationship context), generic CoT is strongest on comparison and inference (where the question logic is not a clean entity chain). Routing dispatches by question type to exploit this complementarity.}

\begin{table}[t]
\centering
\footnotesize
\setlength{\tabcolsep}{2pt}
\caption{\rev{Per-native-type accuracy (\%) of SPARQL CoT (S) and generic CoT (G), no GW, on all three datasets. $\Delta = $ S $-$ G. \texttt{br\_comp} = 2WikiMultiHopQA's \texttt{bridge\_comparison} type.}}
\label{tab:per_type_sparql_vs_generic}
\begin{tabular}{lrrrrrrr}
\toprule
& & \multicolumn{3}{c}{\rev{\textbf{8B}}} & \multicolumn{3}{c}{\rev{\textbf{70B}}} \\
\cmidrule(lr){3-5} \cmidrule(lr){6-8}
\rev{\textbf{Type}} & \rev{\textbf{n}} & \rev{\textbf{S}} & \rev{\textbf{G}} & \rev{\textbf{$\Delta$}} & \rev{\textbf{S}} & \rev{\textbf{G}} & \rev{\textbf{$\Delta$}} \\
\midrule
\multicolumn{8}{c}{\rev{\textit{HotpotQA}}} \\
\midrule
\rev{bridge}     & \rev{398} & \rev{71.1} & \rev{74.9} & \rev{$-3.8$} & \rev{80.2} & \rev{78.4} & \rev{$+1.8$} \\
\rev{comparison} & \rev{102} & \rev{68.6} & \rev{78.4} & \rev{$-9.8$} & \rev{80.4} & \rev{85.3} & \rev{$-4.9$} \\
\midrule
\multicolumn{8}{c}{\rev{\textit{2WikiMultiHopQA}}} \\
\midrule
\rev{compositional}     & \rev{205} & \rev{41.0} & \rev{35.6} & \rev{$+5.4$}  & \rev{48.3} & \rev{46.3} & \rev{$+2.0$}  \\
\rev{br\_comp}& \rev{121} & \rev{37.2} & \rev{46.3} & \rev{$-9.1$}  & \rev{53.7} & \rev{33.9} & \rev{$+19.8$} \\
\rev{comparison}        & \rev{120} & \rev{66.7} & \rev{84.2} & \rev{$-17.5$} & \rev{91.7} & \rev{94.2} & \rev{$-2.5$}  \\
\rev{inference}         & \rev{54}  & \rev{35.2} & \rev{61.1} & \rev{$-25.9$} & \rev{57.4} & \rev{61.1} & \rev{$-3.7$}  \\
\midrule
\multicolumn{8}{c}{\rev{\textit{MuSiQue} (hop-count labels follow native naming)}} \\
\midrule
\rev{2hop}  & \rev{264} & \rev{37.5} & \rev{40.2} & \rev{$-2.7$}  & \rev{49.2} & \rev{41.3} & \rev{$+8.0$}  \\
\rev{3hop1} & \rev{110} & \rev{19.1} & \rev{40.0} & \rev{$-20.9$} & \rev{40.9} & \rev{40.0} & \rev{$+0.9$}  \\
\rev{4hop1} & \rev{60}  & \rev{20.0} & \rev{40.0} & \rev{$-20.0$} & \rev{41.7} & \rev{40.0} & \rev{$+1.7$}  \\
\rev{3hop2} & \rev{36}  & \rev{22.2} & \rev{25.0} & \rev{$-2.8$}  & \rev{38.9} & \rev{27.8} & \rev{$+11.1$} \\
\rev{4hop3} & \rev{17}  & \rev{11.8} & \rev{5.9}  & \rev{$+5.9$}  & \rev{11.8} & \rev{11.8} & \rev{$0.0$}   \\
\rev{4hop2} & \rev{13}  & \rev{15.4} & \rev{7.7}  & \rev{$+7.7$}  & \rev{23.1} & \rev{15.4} & \rev{$+7.7$}  \\
\bottomrule
\end{tabular}
\end{table}

\rev{The complementarity manifests in two ways. On 70B, SPARQL CoT wins on structural types (HotpotQA bridge, 2WikiMultiHopQA compositional and bridge\_comparison with its $+19.8$~pp gain, and every MuSiQue hop-count); generic CoT wins on explicit comparison and inference. On 8B without GW, the SPARQL syntax overhead dominates and generic CoT wins most cells (a deficit GW reverses on bridge-like questions, Table~\ref{tab:routing}). The routing rule (bridge $\to$ SPARQL CoT, comparison and inference $\to$ generic CoT) maps these per-type winners directly, which is why routing is the best of the four methods on every dataset at 70B (Table~\ref{tab:cot_ablation}) and on 2WikiMultiHopQA at 8B.}

\subsection{\rev{Per-Question-Type Gains}}
\label{app:per_type_gains}

\rev{This appendix gives the full per-question-type breakdown of SPARQL CoT gains over baseline (no CoT, no GW) and of GW compression effects, summarised in \S\ref{sec:per_type}.}

\begin{table*}[t]
\centering
\small
\rev{\caption{SPARQL CoT accuracy improvement (pp) over baseline, by question type. HotpotQA: bridge (398) and comparison (102). 2WikiMHQA: compositional (205), bridge\_comparison (121), comparison (120), inference (54). MuSiQue: by hop depth (264/146/90).}}
\label{tab:sparql_type}
\begin{tabular}{l|cc|cccc|ccc}
\toprule
& \multicolumn{2}{c|}{\rev{\textbf{HotpotQA}}} & \multicolumn{4}{c|}{\rev{\textbf{2WikiMHQA}}} & \multicolumn{3}{c}{\rev{\textbf{MuSiQue}}} \\
\rev{\textbf{Model}} & \rev{Bri.} & \rev{Comp.} & \rev{Comp.} & \rev{BrCmp.} & \rev{Cmp.} & \rev{Inf.} & \rev{2h} & \rev{3h} & \rev{4h} \\
\midrule
\rev{8B}  & \rev{$+3.0$} & \rev{$+5.9$} & \rev{$+16.6$} & \rev{$+18.2$} & \rev{$+10.9$} & \rev{$+3.7$} & \rev{$+8.7$} & \rev{$-0.6$} & \rev{$+4.5$} \\
\rev{70B} & \rev{$+2.3$} & \rev{$+2.0$} & \rev{$+4.9$}  & \rev{$\mathbf{+28.1}$} & \rev{$+12.5$} & \rev{$+3.7$} & \rev{$+9.4$} & \rev{$+5.5$} & \rev{$+11.1$} \\
\bottomrule
\end{tabular}
\end{table*}

\rev{The largest gains appear on chain-following types: bridge\_comparison questions see $+18.2$~pp (8B) and $+28.1$~pp (70B) on 2WikiMHQA, while inference questions that require implicit reasoning show the smallest gains ($+3.7$~pp). This aligns with the design of SPARQL CoT: triple-pattern decomposition naturally formalizes entity chains, making it most effective when the question structure matches the formalism. On MuSiQue, the 70B model shows substantial SPARQL CoT benefit at all hop depths ($+9.4$ at 2-hop, $+5.5$ at 3-hop, $+11.1$ at 4-hop), with the largest gain at 4-hop.}

\begin{table}[t]
\centering
\small
\setlength{\tabcolsep}{3pt}
\rev{\caption{GW accuracy $\Delta$ (pp) by question type / hop depth. Counts in parentheses.}}
\label{tab:gw_type}
\begin{tabular}{llrr}
\toprule
\rev{\textbf{Dataset}} & \rev{\textbf{Category}} & \rev{\textbf{SPARQL+8B}} & \rev{\textbf{Base+70B}} \\
\midrule
\multirow{2}{*}{\rev{Hotpot}} & \rev{Bridge (398)} & \rev{$+4.3$} & \rev{$+1.2$} \\
& \rev{Comparison (102)} & \rev{$+12.8$} & \rev{$+1.0$} \\
\cmidrule{1-4}
\multirow{4}{*}{\rev{2Wiki}} & \rev{Composit.\ (205)} & \rev{$+5.8$} & \rev{$-1.0$} \\
& \rev{Br-Comp.\ (121)} & \rev{$+11.6$} & \rev{$+17.4$} \\
& \rev{Comparison (120)} & \rev{$+13.3$} & \rev{$+2.5$} \\
& \rev{Inference\hphantom{00} (54)} & \rev{$+16.7$} & \rev{$+3.7$} \\
\cmidrule{1-4}
\multirow{3}{*}{\rev{MuSiQue}} & \rev{2-hop (264)} & \rev{$-1.9$} & \rev{$-1.2$} \\
& \rev{3-hop (146)} & \rev{$+6.8$} & \rev{$+11.7$} \\
& \rev{4-hop\hphantom{0} (90)} & \rev{$+4.4$} & \rev{$+10.0$} \\
\bottomrule
\end{tabular}
\end{table}

\rev{GW compression similarly benefits deeper hops: on MuSiQue, GW is slightly negative at 2-hop but strongly positive at 3-hop ($+11.7$ for Base+70B) and 4-hop ($+10.0$). On 2WikiMHQA, GW helps most on inference ($+16.7$~pp for SPARQL+8B) and bridge\_comparison questions ($+17.4$~pp for Base+70B), where integrating information across multiple entities benefits most from focused context.}

\subsection{\rev{Compression Baselines}}
\label{app:compression_baselines}

\rev{To assess GW's specific contribution against simpler compression strategies at the same 4k-token budget, we compare it against two baselines on Llama-3.1-8B (where GW's headline effect is most pronounced, Table~\ref{tab:gw_delta}). Results are reported in Table~\ref{tab:compression_baselines}.}

\rev{\paragraph{Baselines.}}
\begin{itemize}
\item \rev{\textbf{Truncation}: take the first $\sim$16k characters of the raw context, cut at the nearest paragraph break in the last 20\% of the budget.}
\item \rev{\textbf{Top-k embedding similarity}: split the context at KET-RAG's natural markers (per-entity, per-relationship, per-text-chunk), embed each chunk with \texttt{all-MiniLM-L6-v2}, embed the question, greedy-fill to budget by cosine similarity, restore original order in the assembled output.}
\end{itemize}

\begin{table}[t]
\centering
\small
\setlength{\tabcolsep}{3pt}
\caption{\rev{Compression-baseline accuracy (\%) on 8B at the same 4k-token budget. \texttt{no-comp} = full retrieved context ($\sim$10k tokens).}}
\label{tab:compression_baselines}
\begin{tabular}{lrrrr}
\toprule
\rev{\textbf{Dataset}} & \rev{\textbf{no-comp}} & \rev{\textbf{+GW}} & \rev{\textbf{+trunc}} & \rev{\textbf{+top-k}} \\
\midrule
\multicolumn{5}{c}{\rev{\textit{Baseline (no CoT)}}} \\
\midrule
\rev{HotpotQA}   & \rev{67.0} & \rev{63.6}          & \rev{34.6} & \rev{65.0} \\
\rev{2WikiMHQA}  & \rev{31.4} & \rev{30.4}          & \rev{18.6} & \rev{30.8} \\
\rev{MuSiQue}    & \rev{23.6} & \rev{19.4}          & \rev{10.2} & \rev{23.6} \\
\midrule
\multicolumn{5}{c}{\rev{\textit{SPARQL CoT}}} \\
\midrule
\rev{HotpotQA}   & \rev{70.6} & \rev{\textbf{76.6}} & \rev{42.4} & \rev{67.4} \\
\rev{2WikiMHQA}  & \rev{45.6} & \rev{\textbf{55.8}} & \rev{32.2} & \rev{47.0} \\
\rev{MuSiQue}    & \rev{28.8} & \rev{\textbf{30.6}} & \rev{17.2} & \rev{27.4} \\
\bottomrule
\end{tabular}
\end{table}

\rev{Three findings emerge. First, truncation collapses uniformly ($-$14 to $-$27~pp vs no-compression on every cell), confirming that compression strategy matters: simply chopping the context is not a valid alternative. Second, top-k embedding similarity matches or slightly beats GW on the no-CoT baseline (HotpotQA $+1.4$, 2WikiMHQA $+0.4$, MuSiQue $+4.2$), but loses to GW on SPARQL CoT across all three datasets ($-9.2$, $-8.8$, $-3.2$). Third, GW's specific contribution is its interaction with SPARQL CoT: the +GW minus top-k gap is largest where structured reasoning has the most leverage (HotpotQA $+9.2$~pp, 2WikiMHQA $+8.8$~pp), consistent with the paper's structural-alignment thesis. GW preserves the entity-relationship chains that triple-pattern reasoning navigates; embedding similarity preserves text similar to the question, not necessarily the bridging structure. We report 8B only since GW's headline effect is on 8B; the 70B comparison is a natural extension.}

\end{document}